\documentstyle[12pt,aas2pp,epsfig]{article}

\def\be{\begin{equation}}
\def\ee{\end{equation}}
\def\ea{{\it et al.}\,}

\def\eg{{\it e.g.},\,}

\def\aa{{\it A\&A}\,}

\def\apjs{{\it ApJS}\,}

\def\rel{relativistic \,}
\def\p{{\rm\ $\pm \;$}}

\begin{document}

\hfill\today

\title{Long RXTE Observations of A2163}

\author{Yoel Rephaeli\altaffilmark{1,2}, Duane Gruber\altaffilmark{3}, 
and Yinon Arieli\altaffilmark{2}}

\affil{$^1$Center for Astrophysics and Space Sciences, University of 
California, San Diego, La Jolla, CA\,92093-0424}

\affil{$^2$School of Physics and Astronomy, Tel Aviv University, 
Tel Aviv, 69978, Israel}

\affil{$^3$4789 Panorama Drive, San Diego CA 92116}

\begin{abstract}

A2163 was observed by the RXTE satellite for $\sim$530 ks 
during a 6 month period starting in August 2004. The cluster 
primary emission is from very hot intracluster gas with 
$kT \sim 15$ keV, but this component does not by itself provide 
the best fitting model. A secondary emission component is 
quite clearly needed, and while this could also be thermal 
at a temperature significantly lower than $kT \sim 15$ keV, 
the best fit (to the combined PCA and HEXTE datasets) is 
obtained with a power law secondary spectral component. The 
deduced parameters of the non-thermal (NT) emission imply a 
significant fractional flux amounting to $\sim 25\%$ of the 
integrated 3-50 keV emission. NT emission is expected given the 
intense level of radio emission, most prominently from a large 
extended (`halo') central region of the cluster. Interpreting 
the deduced NT emission as Compton scattering of the radio-emitting 
relativistic electrons by the CMB, we estimate the volume-averaged 
value of the magnetic field in the extended radio region to be 
$B \sim 0.4 \pm 0.2\, \mu$G.

\end{abstract}


\section{Introduction} 

High quality spatially resolved measurements with current X-ray 
satellites have clearly shown that intracluster (IC) gas is not 
isothermal. With increasing image detail it will likely be found that 
non-isothermality is a general feature of clusters. This is only to 
be expected, given the gas extent, origin, and the processes that have 
affected its evolution. The only justification for spectral fitting of 
the emission from a large cluster region by a single temperature plasma 
emission code (what has been a standard procedure) was insufficient 
spectral and spatial resolution. 

Not only the assumption of isothermality is generally invalid when 
characterizing the emission from a large cluster region, the expectation 
that the emission is {\it purely} thermal may also be doubtful when a 
wide spectral band is considered. This is particularly so in clusters 
with known extended regions of radio emission, where the radio emitting 
relativistic electrons give rise to non-thermal (NT) X-ray emission by 
Compton scattering off the Cosmic Microwave Background (CMB) radiation. 
The well known possibility that cluster X-ray spectra may have high 
energy power law tails (\eg, Rephaeli 1977) has been largely ignored, 
also because of the low sensitivity of past X-ray satellites to the 
detection of the predicted low-level NT emission.

Non-isothermal gas distribution and NT emission are of considerable 
interest both intrinsically, for the detailed understanding of the 
astrophysics of clusters, as well as in the use of cluster properties 
and phenomena as cosmological probes. X-ray emission is currently 
our best probe of the properties of IC gas; the thermal structure of 
the gas can yield important information on energy exchange and 
transport processes. More globally, the detailed gas density and 
temperature profiles are required in analysis of measurements of 
the Sunyaev-Zeldovich (S-Z) effect and its use as a cosmological probe 
of the global parameters of the universe and its large scale structure. 
On the other hand, NT phenomena in clusters have the potential to 
contribute significantly to our understanding of the origin of 
relativistic particles and magnetic fields. It is obvious, therefore, 
that there is strong motivation for a more realistic characterization 
of cluster X-ray spectra.

The search for NT X-ray emission in clusters has been advanced 
considerably by the RXTE and BeppoSAX satellites. We have initiated 
and analyzed long RXTE observations of the Coma cluster (Rephaeli, 
Gruber \& Blanco 1999, Rephaeli \& Gruber 2002), A2319 (Gruber    
\& Rephaeli 2002), and A2256 (Rephaeli \& Gruber 2003). In all 
three clusters we found evidence for a second spectral component. 
While a secondary emission component would generally reflect the 
non-isothermality of IC gas, we found that in these three clusters 
the additional emission is likely to be mostly NT in origin. In each 
of the clusters this conclusion is based on the relative quality of 
the statistical fits (of thermal vs. NT models), and the high physical 
plausibility of NT X-ray emission in the central cluster region in 
which extended radio emission was measured. Similar BeppoSAX searches 
for NT in these and other clusters (including A119, A754, and A2199) 
also yielded evidence for NT emission in at least some of the clusters 
(Fusco-Femiano \ea 1999, Kaastra \ea 1999), Fusco-Femiano \ea 2000, 
Fusco-Femiano \ea 2003). Note that the significance of the BeppoSAX/PDS 
results has been questioned by Rossetti \& Molendi (2003), who claim 
that the level of instrumental error is higher than what was previously 
assumed, and that the detection of NT emission in Coma is much less 
significant than originally reported by Fusco-Femiano \ea (1999). 
However, this claim has been disputed by Fusco-Femiano \ea (2004).

The moderately distant cluster A2163, $z=0.203$, is one of the hottest,  
most luminous clusters, and a prime target of radio, S-Z, and X-ray      
observations in recent years. Its extended central region of radio       
emission (Herbig \& Birkinshaw 1994) was extensively mapped by Feretti 
\ea (2001); this radio `halo' is one of the largest and most luminous. 
Spatial correlation between bright regions in the {\it Chandra} and radio 
maps may be interpreted as evidence for intense merging activity 
(Markevitch \& Vikhlinin 2001), which is thought to enhance the efficiency 
of particle acceleration. The gas temperature profile and a detailed 
temperature map were determined from {\it XMM} (Pratt \ea 2001) and 
{\it Chandra} (Govoni \ea 2004) measurements. Here we report the results 
from a very long observation of A2163, the most distant of the radio 
(`halo') clusters observed with the RXTE.

\section{Observations and Data Reduction}

A2163 Observations with the Proportional Counter Array (PCA) and the 
High Energy X-ray Timing Experiment (HEXTE) on RXTE were made during 
95 separate pointings in the period August 24, 2004 - March 11, 2005. 
Data from PCA detectors 0 and 2 were collected in the `Good Xenon' 
spectral mode, which produces a 256-channel count spectrum nominally 
from 2 to 1000\,keV. The background estimation tool was found to produce 
spectra with a several percent gain error for the epoch of these 
observations; this was corrected to better than 1\% in the software. 
Data from the two independent HEXTE clusters of detectors were taken in 
event-by-event mode, and were subsequently accumulated into 256-channel 
spectra spanning 12--250\,keV. To subtract the background, each HEXTE 
cluster was commanded to beamswitch every $32\,s$ between on-source and 
alternate off-source positions 1.5$^{\circ}$ on either side.

Standard screening criteria were applied to the data segments (Earth
elevation angle, spacecraft pointing, avoidance of the South Atlantic
Anomaly, times of geomagnetic activity), resulting in a net exposure
time for PCA of $528$ ks and an average exposure time of $212$ ks
for the HEXTE clusters.  The HEXTE net observation times were shorter
than the PCA time because HEXTE spends half of the time measuring 
the background, and some more time is lost from electronic dead time 
caused by cosmic rays.

The PCA background was estimated with the `L7/240' faint source model
provided by the instrument team, and was corrected for errors in
detector gain independently in the two detectors. Uncertainties in the
background correction precluded using PCA data above 17 keV. The total 
PCA counting rate was 8 count/s (c/s), or 21\% of background. For HEXTE 
the net counting rate was 0.6 c/s, or 0.4\% of background. The PCA 
background level is estimated to be accurate within $\sim$0.5\%, or 
0.2 c/s. The HEXTE background, which is directly measured, has been 
determined (MacDonald 2000) to be accurate to within a few hundredths 
c/s in long exposures.

\section{Spectral Analysis}

Inspection of the screened light curves for PCA and HEXTE 
revealed no significant variation, as expected for a cluster of
galaxies. Accordingly, we co-added all the selected PCA and HEXTE data 
to form net spectra for analysis. Spectra from the individual PCA
detectors were also coadded, and the small differences in gain were
accounted for in the generation of the energy response matrices, 
following procedures prescribed by the PCA analysis team.

A small energy dependence for systematic errors (e.g., Gruber et al 
2001) averaging about 0.8\%, was applied to the PCA spectrum. 
Additionally, PCA spectral channels below 3\,keV and above 17\,keV 
were excluded because of sensitivity to artifacts in the background 
model, as well as the rapidly declining effective area of the PCA 
outside these bounds. Upon inspection of the high energy channels of 
HEXTE we restricted the analysis to data in the range 12--80\,keV.

We fit the joint PCA and HEXTE spectra summed over all observations to 
three simple spectral models: a Raymond-Smith (R-S) thermal plasma 
emission model, two R-S models at different temperatures, and a R-S 
plus a power law model. In addition we have checked whether we can find 
a meaningful indication for a temperature structure by fitting the 
data with two R-S thermal components plus a power law. In all cases 
most of the observed flux is in a primary $\sim 15$ keV R-S component. 
Best-fit parameters and 90\% confidence intervals are listed in Table 1. 
The best-fit temperature in the single isothermal gas model is 
$15.5 \pm 0.9$ keV, in good agreement with the range determined from 
{\it XMM} (Pratt \ea 2001) and  {\it Chandra} (Govoni \ea 2004). The 
observed $\simeq 6.7$ keV Fe XXV K$_{\alpha}$ line yields an abundance of 
$Z=0.24 \pm 0.13$ $Z_{\odot}$ (where $Z_{\odot}$ denotes solar abundance), 
quite consistent with previously determined values ($\geq 0.2$). No 
cold absorption was measurable, and given the 3 keV PCA threshold, none 
was expected.

The fit to a single isothermal model is only marginally acceptable, 
with $\chi^2 = 43.2$ for 32 degrees of freedom. Residuals have a 
high-low-high pattern which signals the need for another smooth 
spectral component. When a second thermal component is added, best-fit 
parameters are $kT_1 \simeq 20.9$ keV, and $kT_2 \simeq 5.7$ keV, with 
the second component accounting for a minor fraction, roughly 1\% of 
the 3--50 keV flux. For this fit, $\chi^2 = 35.7$, lower by 7.5 than 
the value obtained in the single R-S model. The F-test probability of 
the second component is 0.94 for two additional degrees of freedom. 
Although the $\chi^2$ is acceptable for this fit, the 90\% error 
bounds of [16,80] keV for $kT_1$ lie outside - in fact, well 
outside - the more tightly constrained values of $11.5 \pm 1.5$ keV 
obtained with {\it ASCA} (Markevitch 1996), $14.6 \pm 0.9$ keV with 
{\it ROSAT} (Elbaz et al. 1994), $14.6 \pm 0.5$ keV with {\it XMM} (Pratt 
\ea 2001), and $12.4\pm 0.7$ keV measured with {\it Chandra} (Govoni 
et al. 2004). We therefore view this solution as doubtful. Indeed, 
the range of solutions is very wide because the problem is numerically 
highly degenerate. If we consider the possible sets of ($kT_1$, $kT_2$), 
the value $\chi^2 + 4.6$ defines the joint 90\% probability contour 
in parameter space. The lowest temperature combination permitted in 
this range is $kT_1 = 16.1$ keV and $kT_2 = 0.8$ keV, with fractions 
of the 3--50 keV flux, respectively, of 99\% and 1\%. The highest 
temperature combination has $kT_1 \simeq 14.3$ keV and $kT_2$ 
unbounded, with 13\% of the 3--50 keV flux in the higher temperature 
component. These results seem to indicate that the two temperature 
fit improves on a single temperature one mostly by accounting for 
emission at the higher energy channels.

When the second component is a power law the best-fit has $\chi^2= 
32.2$, fully 11.0 lower than for the isothermal fit. The best-fit 
power law index is $\alpha =1.8$ with 90\% confidence bounds of 
[-2.4, 2.7], and the 3-80 keV flux is $2.2\cdot 10^{-11}$ erg/(cm$^2$ s), 
with 90\% confidence bounds of $(0.3 - 4.1)\cdot 10^{-11}$ erg/(cm$^2$ s). 
The F-test probability of this power law component is 0.99 for two 
additional degrees of freedom. The 3--50 keV flux of the power law 
component is 27\% of the total. This component comprises most of the 
flux at energies $\geq 50$ keV. The iron abundance is $Z=0.32Z_{\odot}$, 
with 90\% confidence bounds of $[0.11, 0.79]Z_{\odot}$. The best-fit 
fluxes from the combined thermal and power law emissions, and that of 
just the power law component, are shown in Figure 1, together with the 
data points. Residuals of this best fit model are plotted in the lowest 
panel; for comparison, the residuals to the single temperature model are 
shown in the middle panel. Differences between the latter two plots are 
apparent mostly at high energies.

The results of the temperature mapping with {\it XMM} (Pratt \ea 2001) 
and {\it Chandra} (Govoni \ea 2004) clearly necessitate inclusion of a 
temperature distribution in the fits. The RXTE FOV includes a region 
which is much larger than that probed by both {\it XMM} and {\it Chandra}. 
This, and the fact that the RXTE lacks spatial resolution, make the 
RXTE insensitive to even large scale features in the {\it Chandra} 
temperature map. The above fit to a two-temperature model grossly 
samples substantially different temperature components. Just so as to 
verify that we cannot get additional useful information from a more 
detailed spectral modeling, we have considered also a model with 
three components - two R-S and a power law. When the model parameters 
are fully unconstrained their values are obviously very loosely 
determined: The best-fit to this model - with $\chi^2=31.1$ for 28 dof - 
yields $kT_1 = 16.3 \pm 1.2$ keV, a negligible second thermal component 
with $kT_2 = 3.7 \pm 7.5$ keV, $Z=(0.28 \pm 0.16)Z_{\odot}$, $\alpha = 1.3 
\pm 3.0$, and power law normalization of $(1.3 \pm 2.8) \cdot 10^{-11}$ 
erg/(cm$^2$ s). The decrease in $\chi^2$ when a power law component is 
added is $4.6$, so (from F-test statistics) the significance is about 
85\% (or, more specifically, 83\% - 86\%, depending on the particular 
choice of `interesting' parameters). However, the $\chi^2$ hypersurface 
around the minimum correspoding to this parameter set is not very deep, 
meaning that a small increase of $\chi^2$ above its minimum value results 
in a rather different parameter set. In spite of the implied substantial 
uncertainty, we find that the need for a power law component is 
removed only when $kT_1 \geq 30$ keV, which we consider to be rather 
unrealistic.

A power law component at the above deduced level is apparent in the 
HEXTE measurements. To check general consistency between results from 
the PCA and HEXTE, we have determined the parameters of this component 
by analyzing HEXTE data separately. Clearly, due to the relatively high 
(12 keV) lowest energy channel of HEXTE, an appreciable fraction of the 
thermal emission cannot be detected by this experiment, so we have set 
the parameters of the isothermal component to its best-fit values from 
the joint (PCA \& HEXTE) data. Fitting for the power law, we obtain 
$\alpha \simeq 1.9_{-0.6}^{+0.9}$, and $F(3-80 \,keV) 
\simeq (2.5_{-0.5}^{+1.2})\times 10^{-11}$ erg/(cm$^{2}$ s) for 
the power law (photon) index and the integrated 3-80 keV flux. 
Thus, HEXTE data alone provide significant additional evidence for NT 
emission whose spectral index has a narrower 90\% confidence interval 
than that determined in the joint fits. While this is a constrained 
fit, with only two free parameters, it is both physically (by virtue 
of the anticipated levels of thermal and NT emissions) and practically 
(given the relative insensitivity of HEXTE to emission at low energies) 
well motivated.

\begin{table*}
\caption{Results of the spectral analysis}
\vspace{0.5cm}

\begin{tabular}{|l|ccc|}
\hline
Parameter        & single R-S        & two R-S       &  R-S$+$power law \\
\hline
$\chi^2/dof$       & 43.2/32          & 35.6/30       & 32.2/30          \\
$kT_1$ (\rm{keV}) & 15.5 [14.6, 16.4] & 20.9 [15.6, 80] & 14.5[12.1, 21.0] \\
\,\, Normalization$^a$  &  0.450 [0.043, 0.047] &  0.036 [0.001, 0.044] & 
0.036 [0.022, 0.046] \\ 
Fe abundance$^b$     & 0.24 [0.12, 0.37]  & 0.27 [0.12, 0.54] &  0.32 [0.11, 
0.79] \\
$kT_2$ (\rm{keV})  &                &  5.7 [0.8, 14.3]   &    \\
\,\, Normalization$^a$  &           & 0.014 [0.001, 0.045] &  0.0035   \\
$I_\epsilon$(5\, \rm{keV})
(\rm{$cm^{-2}\,s^{-1}$})    &      &    & (1.33$\pm 0.45)\times 10^{-4}$  \\
Photon index      &                  &                & 1.8 [-2.4, 2.7]    \\
\hline
\end{tabular}

\tablenotetext{}{Notes: }

\tablenotetext{}{All quoted errors are at the 90\% confidence  level. }

\tablenotetext{a}{e.m. = Raymond-Smith emission measure in units of
10$^{-14} \int N_eN_H dV$ / 4$\pi D^2$, where $D$ is the luminosity 
distance and $N_e$, $N_H$ are the total number of electrons and 
protons, respectively.}

\tablenotetext{b}{Abundance is expressed relative to solar values.}

\end{table*}

\begin{figure}
\hspace{-1.5cm}
{\psfig{file=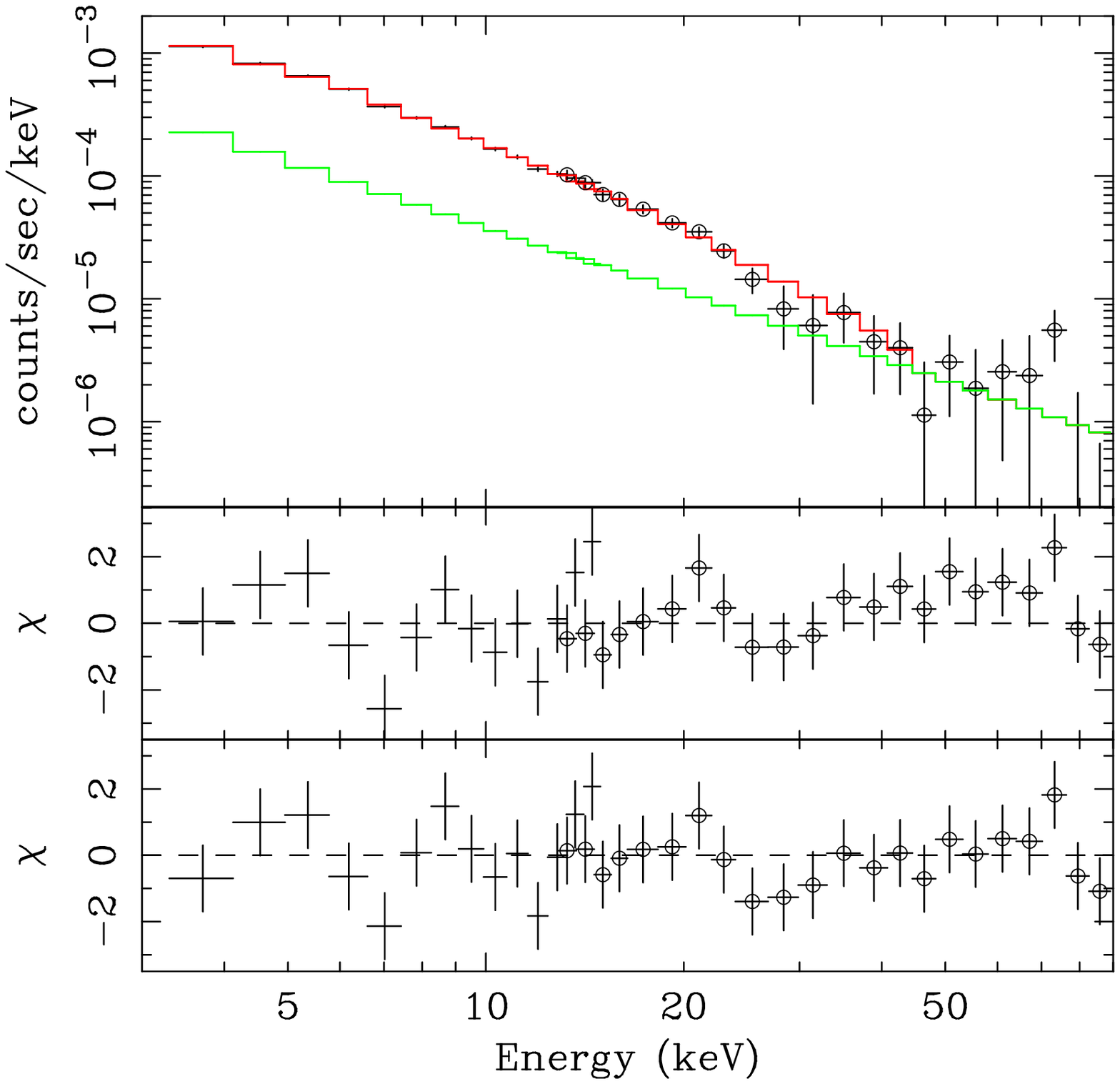, width=9cm, angle=270}}
\figcaption{The RXTE (photon) spectrum of A2163 and folded 
Raymond-Smith ($kT_1 \simeq 14.5$ keV) and power law (index $\alpha = 
1.8$) models. HEXTE data points are marked with circles and 68\% error 
bars. The total fitted spectrum is shown with a histogram, while the lower 
histogram shows the power law portion of the best fit. The quality of 
the fit is demonstrated in the lowest panel, which displays the observed 
difference normalized to the standard error of the data point. For 
comparison, the residuals of the fit to a single temperature model are 
shown in the middle panel. The improved quality of the fit to the 
combined thermal plus power law components is apparent mostly at high 
energies.}
\end{figure}

We note that a relatively short $\sim 109$ ks observation of A2163 with 
BeppoSAX/PDS yielded an upper limit on power law emission, 
$F(20-80 \, keV) \leq 5.6\times 10^{-12}$ erg/(cm$^{2}$ s) 
(Feretti \ea 2001). Our much deeper observation yields a significant 
detection of NT emission with a flux level $F(20-80 \, keV) 
\simeq 1.1_{-0.9}^{+1.7}\times 10^{-11}$ erg/(cm$^{2}$ s), 
whose 90\% confidence interval includes the PDS upper limit.

\section{Discussion}

The results of the above spectral analysis clearly indicate that 
the emission in A2163 cannot be described by a single temperature 
emission model (as is apparent from positive residuals at both 
low and high energies). A better fit is provided by a two-temperature 
model, with the secondary thermal component accounting for the low 
energy residuals. The {\it Chandra} temperature map of the central 
cluster region shows a region, roughly $2'X2'$ in area, where the 
temperature is $\sim 9$ keV. The secondary thermal component we deduce 
has a somewhat lower value, but given the large uncertainty in the 
value we deduce, the difference is not significant, as is the 
{\it rough} - due to the substantially larger ($58'$ FWHM) 
RXTE FOV, which includes the emission from the cooler outer regions 
- consistency of the relative flux values of the two thermal components. 

However, in the best fitting model the second component is not thermal, 
but rather a NT power law. Indeed, NT emission is expected from the 
populations of \rel electrons that emit both the extended IC radio 
emission, as well as the emission from the several powerful radio 
sources in the central $\sim 2$ Mpc region of A2163. In Table 2 we 
list the measured radio fluxes and spectral (energy) indices for the 
dominant sources in the RXTE FOV. In addition to the remarkably 
luminous emission from the most extended, regularly-shaped `halo' 
region, we include the emission from four sources measured by Feretti 
et al. (2001). Fluxes and spectral indices for the `halo' and 'relic' 
sources are determined from VLA measurements at 6 \& 20 cm. To 
estimate the expected X-ray fluxes also from the three tailed (T) 
sources, we need to specify their spectral indices; since these were 
not given by Feretti \ea, we adopt a typical value of $0.8\pm 0.1$.

\begin{table*}  
\caption{Values of radio parameters (and 1$\sigma$ errors)}
\vspace{0.5cm}
\small
\vspace{0.5cm}
\begin{tabular}{ l@{\hspace{0.3cm}} c@{\hspace{0.4cm}}
    c@{\hspace{0.4cm}} c@{\hspace{0.4cm}} } \hline
Source & Flux Density - 20 cm (mJy) & Spectral (energy) index & Size (kpc)\\
\cline{1-4}
Halo & 155\p2 & 1.6\p0.3 & 2070\p70 \\
Relic & 18.7\p 0.3  & 2.1\p0.3 &460\p40 \\
J1615-062 (T1) & 34.5\p0.5 & 0.8\p 0.1 & 450 \\
J1615-061 (T2) & 6.0\p0.3 & 0.8\p 0.1 & 120 \\
J1616-062 (T3) & 24.9\p0.4 & 0.8\p 0.1 &270 \\
\cline{1-4}
\end{tabular}
\tablenotetext{}{Source size is based on a Hubble constant of 70 
$km\,s^{-1}\,Mpc^{-1}$}

\end{table*}

On the relevant spatial scales the radiation field most tightly 
coupled to the electrons by Compton scattering is the CMB, coupling 
that is only further enhanced in A2163 ($z=0.203$) due to the 
$(1+z)^4$ dependence of the CMB energy density. In order to quantify 
the emission that effectively results from the scattering, we need 
to determine the spectral density distribution of the electrons. 
Radio emission from the central extended source dominates the overall 
emission, and since we expect the mean, volume-averaged field in this 
extended source to be much lower than field strengths in the galactic 
radio sources, it is clear that most of the Compton-produced X-ray 
emission comes from the 'halo'. Nonetheless, we have estimated the 
emission by \rel electrons in the other sources listed in Table 2.

Compton fluxes of the radio sources in Table 2 can be estimated from 
their measured radio fluxes and estimated mean field values. The 
latter can be determined by assuming energy equipartition between 
particles and fields. This assumption may be roughly valid in the 
particles galactic sources, by virtue of the relatively short 
timescales of all the relevant processes governing the particles 
and fields - acceleration, couplings to the field and interstellar 
gas. (We note in passing that the attainment of equipartition in the 
IC space is questionable: IC fields and particles are likely to 
be of galactic origin, but the evolution during their expulsion 
from galaxies and in IC space is quite different; effective 
equilibration is not likely to occur there.) Using the deduced 
values of the equipartition in the radio sources we have estimated 
their respective Compton X-ray fluxes. These turn out to be quite 
small for the relic as well as the three tailed sources, whose 
combined flux in the 3-80 keV band is $\leq 1\%$ of the measured 
value. (Note that consideration of the dominant non-'halo' radio 
sources in the clusters in which we have previously found evidence 
for NT emission - Coma, A2256, and A2319 - similarly shows that 
their predicted X-ray emission contributes negligibly to the 
deduced NT emission in each of these clusters.)

Since the known radio sources in the central region of A2163 do not 
contribute appreciable X-ray emission, we attribute all the measured 
emission to electrons in the 'halo', and proceed to determine the mean, 
volume-averaged field in this region, $B_{rx}$, using the 
Compton-synchrotron formulae (\eg Rephaeli 1979). For consistency 
with the basic premise that most of the measured Compton flux is 
due to electrons in the `halo', and the fact that the electron 
spectrum can be more precisely deduced from the radio data, we use 
the radio spectral index in order to infer the electron index. Doing 
so we compute $B_{rx}\simeq 0.4\pm 0.2 \,\mu$G. It should be emphasized 
that the overall uncertainty in the estimated mean field is substantially 
higher than this formal error due to the lack of spatial information 
on NT X-ray emission, which necessitates invoking the assumption that 
the emitting \rel electrons and the field are co-spatial. 

As a consequence of the assumption that the spatial factors in the 
theoretical expressions for the radio and NT X-ray fluxes are roughly 
equal, it follows that the mean value of the deduced magnetic field is 
independent of the source size and distance. The \rel electron energy 
density does depend on these parameters. Scaling to the observed radius 
of the diffuse radio emission, and integrating the electron spectral 
distribution over all energies above 1 GeV, we obtain $\rho_e \simeq 
1.4_{-0.7}^{+0.8} \times 10^{-13} h_{70}^{-1} \,erg/cm^3$, where 
$h_{70}$ is the value of the Hubble constant in units of 
$70$ km\,s$^{-1}$\,Mpc$^{-1}$). Based on the high Galactic proton to 
electron energy density ratio of cosmic rays, it can be conjectured 
that the energetic proton energy density much higher than this 
value. A more specific assessment cannot be made without consideration 
of the electron and proton origin and their respective energy losses. 

IC magnetic fields can also be estimated from Faraday rotation 
measurements of background radio sources seen through clusters, 
yielding a different mean field value, $B_{fr}$. These measurements 
usually yield field values that are a few $\mu$G (see, \eg, Clarke, 
Kronberg, and B\"ohringer 2001, and the review by Carilli \& Taylor 
2002), up to an order of magnitude higher than values of $B_{rx}$. 
Average field strength of a few $\mu$G over an extended cluster region 
would have important implications on the range of electron energies 
that are deduced from radio measurements, and therefore on the electron 
(synchrotron and Compton) energy loss times. Weaker mean 
(volume-averaged) fields imply higher electron energies, and therefore 
shorter electron energy loss times, with possibly important ramifications 
for \rel electron models (\eg, Rephaeli 1979, Sarazin 1999, Ensslin \ea 
1999, Brunetti \ea 2001, Petrosian 2001). However, the apparent discrepancy 
between deduced values of $B_{rx}$ and $B_{fr}$ is usually naively 
interpreted. $B_{rx}$ and $B_{fr}$ actually provide very different 
measures of the mean field strength: $B_{rx}$ is essentially a weighted 
volume-average of the \rel electron density and of $B^q$, with $q \geq 2$; 
on the other hand, $B_{fr}$ is an average of the product of the line of 
sight component of the field and the gas density. In general, the overall 
spatial dependence of these averages are considerably different, and so 
are their deduced values. Specific examples and quantitative comparisons, 
including also the impact of observational uncertainties, are given by 
Goldshmidt \& Rephaeli (1993), and by Newman, Newman \& Rephaeli (2002).

\acknowledgments

We are grateful to the referee for very helpful critical comments. 
This work was supported by a NASA grant at UCSD.

\parskip=0.02in
\def\ref{\par\noindent\hangindent 20pt}

\end{document}